\documentclass[10pt]{elsarticle}
\usepackage{epsfig}
\usepackage{amssymb}
\usepackage{graphicx}
\usepackage{subfigure}
\usepackage{rotating}
\journal{New Astronomy}
\newcommand{\de}{{\rm d}}
\newcommand{\bea}{\begin{eqnarray}}
\newcommand{\eea}{\end{eqnarray}}

\begin{document}
\begin{frontmatter}

\title{Star formation in high redshift galaxies including Supernova feedback:
effect on stellar mass and luminosity functions}
\author{Saumyadip Samui\corref{cor1}}
\ead{ssamui@gmail.com}
\cortext[cor1]{Department of Physics, Presidency University, 86/1 College
Street, Kolkata - 700073, India.}
\address{Department of Physics, Presidency University, 86/1 College Street,
Kolkata - 700073, India.}

\begin{abstract}
We present a semi-analytical model of high redshift galaxy formation. In our
model the star formation inside a galaxy is regulated by the feedback from
supernova (SNe) driven outflows. We derive a closed analytical form for star
formation rate in a single galaxy taking account of the SNe feedback in
a self-consistent manner. We show that our model can explain the observed
correlation between the stellar mass and the circular velocity of galaxies
from dwarf galaxies to massive galaxies of $10^{12}~M_\odot$. For small mass
dwarf galaxies additional feedback other than supernova feedback is needed to
explain the spread in the observational data. Our models reproduce the observed
3-D fundamental correlation between the stellar mass, gas phase metallicity and
star formation rate in galaxies establishing that the SNe feedback plays a
major role in building this relation. Further, the observed UV luminosity
functions of Lyman-Break galaxies (LBGs) are well explained by our feedback
induced star formation model for a vast redshift range of $1.5 \le z \le 8$. 
In particular, the flattening of the luminosity functions at the low luminosity
end naturally arises due to our explicit SNe feedback treatment.

\end{abstract}
\begin{keyword}
galaxies: high-redshift;
galaxies: star formation;
stars: winds, outflows;
galaxies: luminosity function;
\end{keyword}
\end{frontmatter}

\section{Introduction}
Presently we possess a wealth of observations regarding the high redshift
universe, thanks to present day technology. The galaxies are regularly
being detected till redshift $z \sim 10$ using Lyman Break technique (Steidel
et al., 2003). The UV luminosity functions of Lyman break galaxy (LBG) are well
constrained upto redshift $z \sim8$ using Hubble Ultra deep field observations
(Bouwens et al., 2007; Bouwens et al. 2008; Reddy \& Steidel, 2009; Oesch et
al., 2010; Bouwens et al., 2011). The faint end slope of the UV luminosity
functions is well established for $z\lesssim 6$ and it shows flattening of the
luminosity function at low luminosity end. The presence of Gunn-Peterson (Gunn
\& Peterson, 1965) absorption in the spectrum of high redshift quasars tells us
a transition from highly ionised inter galactic medium (IGM) to partially
ionised IGM around redshift $z \sim 6$ (Wyithe, Loeb \& Carilli, 2005; Fan et
al., 2006; Mortlock et al., 2011). Also the electron scattering optical depth
($\tau_e$) measured in the Cosmic Microwave Background Radiation (CMBR) by
Wilkinson Microwave Anisotropy Probe (WMAP) constrains the reionization
redshift to be $z_{re}=10.4\pm1.2$ (Komatsu et al. 2011) for a step
reionization scenario.

Further, the high resolution absorption spectra of quasars show the presence
of metals in very low density IGM far away from galaxies. Metals are produced
inside galaxies and believed to be transported by outflows produced by the
Supernova (SNe) explosions in the galaxy. The outflows are routinely being
observed in low redshift galaxies as well as in high redshift galaxies (Martin,
1999; Pettini et al., 2001; Martin, 2005). These outflows are likely to
expel metals along with a large amount of inter stellar medium (ISM).
This, in turn, reduces the star formation in galaxies by reducing the available
gas to form new generation of stars. Thus the supernovae give a negative
feedback to the star formation by throwing out gas from galaxies in the form
of galactic winds.

Not only that, outflows also regulate the amount of metals in galaxies.
Starting from very early days, a tight correlation has been observed
between luminosity and metallicity of galaxies (Lequeux et al. 1979).
Later, a more fundamental correlation has been found between the
stellar mass and gas phase metallicity of galaxies in the local
universe (Garnett, 2002; Tremonti et al., 2004; Lee et al., 2006;
Kewley \& Ellison, 2008) as well as in high redshift galaxies
(Savaglio et al., 2005; Erb et al., 2006; Mannucci et al., 2009;
Mannucci et al., 2010). It has been observed that galaxies with higher stellar
mass trend to have more metals compared to their lower stellar mass counter
parts. Further observations of local as well as high redshift universe
show that the luminosity-metallicity or mass metallicity relation observed in
galaxies is due to a more general relationship between stellar mass,
metallicity of the gas and star formation rate (Mannucci et al. 2010;
Cullen et al. 2013). Since, the outflows from galaxies throw metal enrich gas
it is most likely that outflows play an important role in building up this
correlation (Kobayashi et al., 2007; Scannapieco et al., 2008).

It has been seen from observations that the amount of gas/ISM expelled from
a galaxy due to outflows is inversely proportional to the mass of the galaxy
(Martin, 1999; Martin 2005). This is expected if the hot gas produced in SNe
explosions drives the outflow. Roughly 10\% of total SNe energy would be
available for driving the outflow when the SNe remnants started to overlap with
each other (Cox, 1972). The conservation of SNe energy available to drive the
outflow to the kinetic energy of outflowing gas would lead to a mass outflow
rate inversely proportional to the square of the circular velocity ($v_c$)
of the galaxy. Even if the hot gas loses its thermal energy due to
radiative cooling the momentum of the gas and/or the cosmic rays
produced in the SNe shocks can still drive the outflow (Samui et al., 2010;
Ostriker \& Mckee 1988). In such cases as well the inverse relation
between outflowing mass and the circular velocity of the galaxy still
holds with a different scaling. Thus due to supernova explosions
small mass galaxies would lose more gas and experience a strong
negative feedback to the star formation compare to higher mass galaxies.

Hence, it is important to build a complete model of high redshift galaxy
formation taking account of all the observational evidences, particularly
the SNe feedback driven star formation in high redshift galaxies
and their metal transport to the IGM. Numerical simulations are the
best way to study all these together and a tremendous effort is going on
(for example, Scannapieco et al., 2005 \& 2006; Dave, Oppenheimer
\& Sivanandam, 2008; Dave, Oppenheimer, \& Finlator,
2011; Scannapieco et al., 2012). However, present state of art
hydrodynamic simulations are far from reality. They are constrained by the
resolutions as well as the amount of physical processes that they can
take account together (Scannapieco et al., 2012; Stringer et al. 2012).
Here, we build an analytical model of star formation in the high redshift
galaxies regulated by the feedback from SNe driven winds and try to explain
the amount stellar mass and metals detected inside galaxies and the high
redshift UV luminosity functions of Lyman-Break Galaxies (LBGs).
In past, several authors have proposed semi-analytical models in order
to understand the high redshift as well as low redshift galaxy formation
process (White \& Frenk, 1991; Kauffmann, White \& Guiderdoni, 1993; Cole
et al., 1994; Baugh et al, 1998; Somerville \& Primack, 1999; Chiu \& Ostriker,
2000; Granato et al 2000; Choudhury \& Srianand, 2002; Baugh et al., 2005;
Shankar et al 2006). These works clearly demonstrated the power
of such semi-analytical modeling by predicting various observations
regarding high redshift universe. However, they have not derived a universal
closed analytical form for the time evolution of star formation rate (SFR)
in a single galaxy. Chiu \& Ostriker (2000), Choudhury \& Srianand (2002)
and some of our earlier works (Samui et al., 2007, 2008; Jose et al., 2011)
have used such a closed form without considering the SNe feedback and also
not deriving their star formation model from first principle.
Others have just considered star formation rate to be proportional to the
available cold gas and not derived a single evolution equation
for the evolution of the star formation rate including feedback.
Here, we solve for the star formation rate in a closed form starting from
very basic physics governing the star formation and taking account of the
negative feedback from galactic outflows on the star formation in a
self-consistent manner. This closed form will be very useful while fitting
the photometric observations of high redshift galaxies in order to find their
star formation history, stellar mass etc. Moreover, the new data are extended
to much lower in stellar mass/luminosity and higher redshift where the process
of reionisation is still going on. In one hand these low mass systems are the
dominating sources of reionization. On the other hand they are much likely to
prone to SNe feedback. Hence, it is timely to revisit feedback induced star
formation in high redshift galaxies in the light of new improved data sets.
Further, semi-analytical models are always useful as they are computationally
inexpensive and help to understand the average universe very well. Also it is
important to explore vast range of parameters that regulates the physical
processes happening inside a galaxy. 

The paper is organised as follows. In section~\ref{sec_sfig} we clearly state
our feedback induced star formation model in galaxies and how well it explains
the stellar mass detected in dwarf galaxies to high mass galaxies. The
mass-metallicity-SFR relation of galaxies is discussed in section~\ref{sec_mass_metal}.
We present our model of UV luminosity function in section~\ref{sec_lf}.
We show our model predictions of UV luminosity functions of LBGs and compare
that with observations in section~\ref{sec_fit}. Finally we draw our
conclusions with some discussions in section~\ref{sec_cd}. Through out this
paper we assume a $\Lambda$cold dark matter ($\Lambda$CDM) cosmology with the
cosmological parameter as obtained by recent WMAP observation\footnote{For
a list of Cosmological Parameters based on the latest observations see
http://lambda.gsfc.nasa.gov/product/map/current/parameters.cfm}, i.e.
$\Omega_\Lambda=0.73$, $\Omega_m = 0.27$, $\Omega_b=0.045$ and Hubble parameter
$H_0=70$~km/s/Mpc.

\section{Feedback induced Star formation in individual galaxy}
\label{sec_sfig}
We model the star formation rate including the feedback from SNe driven
outflows in a galaxy as follows. We assume that the instantaneous star
formation rate at a given time is proportional to the amount of cold gas
present in the galaxy. Once the dark matter halo virialises
it accretes baryonic matter and a fraction, $f_*$, of that becomes
cold and available for star formation. The $f_*$ can be thought of as
star formation efficiency and is a free parameter in our model.
The baryon accretion rate in a galaxy of total mass $M$ at time $t$ after
the formation of dark matter halo is taken as
\begin{equation}
\frac{dM_g}{dt}=\left(\frac{M_b}{\tau}\right)e^{[-t/\tau]}
\label{eqn_accretion}
\end{equation}
where $M_g$ is the gas mass and $M_b = (\Omega_b/\Omega_m) M$ is the
total baryonic mass in the halo. Further, $\tau$ is the dynamical time
of the galaxy (Barkana \& Loeb, 2001). We assume that the total gas mass
in the galaxy is equal to the dark matter mass times the universal 
dark matter to baryon mass ratio. Note that integrating Eq.~\ref{eqn_accretion}
from $t=0$ to $\infty$ results $M_g=M_b$. 
The exponential form of the baryon accretion rate can be understood
as follows. Once the dark matter halo virialises, the baryons are captured
in the potential well and heated to virial temperature of the dark matter
potential. In order to form stars the gas needs to cool and fall
into the centre of the galaxy. If one assumes the rate of hot gas becoming
cold is proportional to the amount of hot gas present, the increase in
cold gas mass would follow an exponential form with time scale
governed by the cooling time ($t_{\rm cool}$). However, as already mentioned,
this cold gas has to collapse into the centre of the galaxy in order to form
stars. The collapse of the gas into the centre of dark matter halo is governed
by the dynamical time scale of the gravitational potential. For most of the
galaxy masses that we are interested, $t_{\rm cool} \lesssim \tau$ (Silk, 1993).
This is true even at the mean overdensity of the collapsed halo which is
$\sim$~180 times the background density. Hence the effective cold gas accretion
rate can be taken as an exponential form with time scale of the order of
dynamical time scale for the system. Further note that the accretions of cold
gas in galaxies has been found to be important observationally and also
considered in previous semi-analytical models (i.e. Kauffmann et al. 1993,
Shankar et al. 2006). Our exponential cold gas accretion rate tries to model
that. Such an exponential form for the cold gas accretion rate has been also
used by other semi-analytical works (for example see appendix of Shankar et al.
2006). 

At a given time some gas would already be locked inside stars. Further, the
massive stars are short lived and would explode as supernova after few times
$10^7$~yrs. These supernovae can drive the cold gas out of galaxy as galactic
wind. We assume that the amount of gas mass driven out by the supernova in the
form of wind is proportional to the instantaneous star formation rate,
neglecting the time delay of $\lesssim 10^7$~yrs between the star formation and
subsequent explosion of supernova, i.e. 
\begin{equation}
\dot{M}_w = \eta_w \dot{M}_{*}.
\label{eqn_eta}
\end{equation}
Here, $M_w$ is the wind mass; $M_*$ is the star mass and over dot represents the
time derivative. The proportionality constant $\eta_w$ depends on the driving
mechanism of outflows. For example, if the outflows are driven by the
hot gas produced by the SNe, then $\eta_w \propto v_c^{-2}$, $v_c$ being
the circular velocity of the galaxy. Such outflows are referred to as
energy driven outflows. If the hot gas loses its energy by radiative cooling,
then outflows can be potentially driven by the pressure of cosmic rays produced
in the SNe shocks. It was shown analytically in Samui et al. (2010) that in
such cases $\eta_w \propto v_c^{-2}$ as well. However, if the momentum of the
cooled gas helps in driving the outflow, $\eta_w \propto v_c^{-1}$. These are
called momentum driven outflows. Detailed models of such supernova driven
outflows can be found in Weaver et al. (1977), Ostriker \& McKee (1988),
Scannapieco et al. (2002), Veilleux, Cecil, \& Bland-Hawthorn (2005), Samui
et al. (2008).

Finally, taking account of the mass in stars, mass in outflows and baryonic
mass accreted we can write down the star formation rate at time $t$ after
the formation of dark matter halo as
\begin{equation}
\frac{d M_*}{dt} = f_t \left[\frac{ f_* M_g - M_* - M_w }{ \tau }\right],
\label{eqn_dmdt}
\end{equation}
with $f_t$ is some proportionality constant that governs the duration of
star formation. Taking the time derivative of Eq.~\ref{eqn_dmdt} and putting
values of $\dot{M}_g$ and $\dot{M}_w$ from Eq.~\ref{eqn_accretion} and
\ref{eqn_eta} respectively, we get
\begin{eqnarray}
\frac{d^2M_*}{dt^2}&=&\frac{f_t}{\tau}\left[f_*\frac{dM_g}{dt} - \frac{d M_*}{dt} - \frac{dM_w}{dt}\right] \nonumber \\
&=&\frac{f_t}{\tau}\left[f_*\frac{M_b}{\tau}e^{-\frac{t}{\tau}} - \frac{d M_*}{dt} - \eta_w\frac{dM_*}{dt}\right].
\end{eqnarray}
Integrating this with the boundary condition that $dM_*/dt=0$ and $M_*=0$ at
$t=0$ we have
\begin{equation}
\frac{d M_*}{dt} = \frac{M_b f_* f_t}{\tau [f_t(1+\eta_w)-1)]}\Bigg[ e^{-\frac{t}{\tau}} - e^{-f_t(1+\eta_w)\frac{t}{\tau}} \Bigg].
\label{eqn_dM_dt_2}
\end{equation}
Eq.~\ref{eqn_dM_dt_2} gives the analytical form of star formation rate in a
galaxy at age $t$ in presence of supernova feedback and is the backbone of
our feedback induced star formation model. Note that putting $\eta_w$ equals
to zero would lead to star formation rate in a without feedback scenario and
that form of star formation rate has been widely used in semi-analytical galaxy
formation model in past by various authors (see Chiu \& Ostriker, 2000;
Choudhury \& Srianand, 2002; Samui et al., 2007; Jose et al., 2011; Jose et
al., 2013).

It is interesting to note that the total baryonic mass that will be
eventually converted to stars is
\begin{equation}
M_*(t \to \infty) = \left(\frac{1}{1+\eta_w}\right) M_b f_*
\label{eqn_m_total}
\end{equation}
which is independent of $f_t$ and inversely proportional to $1 + \eta_w$.

We have already mentioned that
\begin{equation}
\eta_w=\left(\frac{v_c}{v_c^0}\right)^{-\alpha},
\label{eqn_eta_w}
\end{equation}
where  $v_c^0$ is the circular velocity of the halo for which $\eta_w=1$
and $\alpha = 2$ or $1$ depending on whether outflows are energy driven/cosmic
rays driven or momentum driven. Hence, for a small mass galaxy that has
a lower circular velocity, $\eta_w$ is large compare to a high mass galaxy
with higher $v_c$. Thus from Eq.~\ref{eqn_m_total} it is clear that our model
predicts a higher stellar mass fraction ($M_*/M_b$ or equivalently $M_*/M$)
in high mass galaxies and lower stellar mass fraction for
small mass dwarf galaxies. The SNe feedback regulates the total amount of
star formed in a galaxy. For small mass dwarf galaxies the SNe feedback is
strong leading to a large decrease in the star formation in those galaxies.
Large mass galaxies due to their large gravitational potential are less prone to
the SNe feedback and suppression of star formation due to such feedback is small.
Indeed, observationally we found such a correlation
between the stellar mass and circular velocity/mass of galaxy.

\begin{figure*}
\centerline{\epsfig{figure=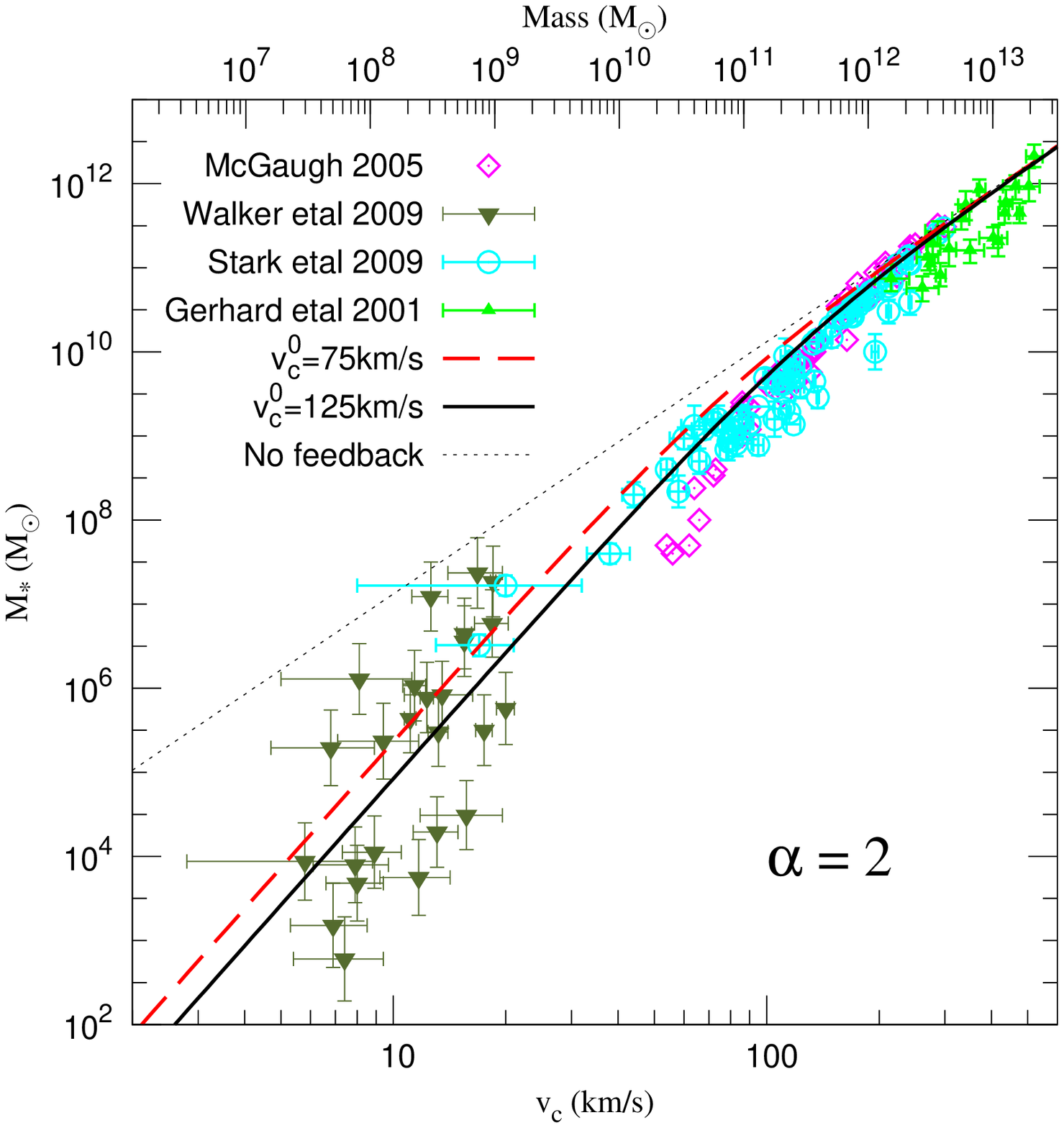,width=7.5cm,angle=-90.}\epsfig{figure=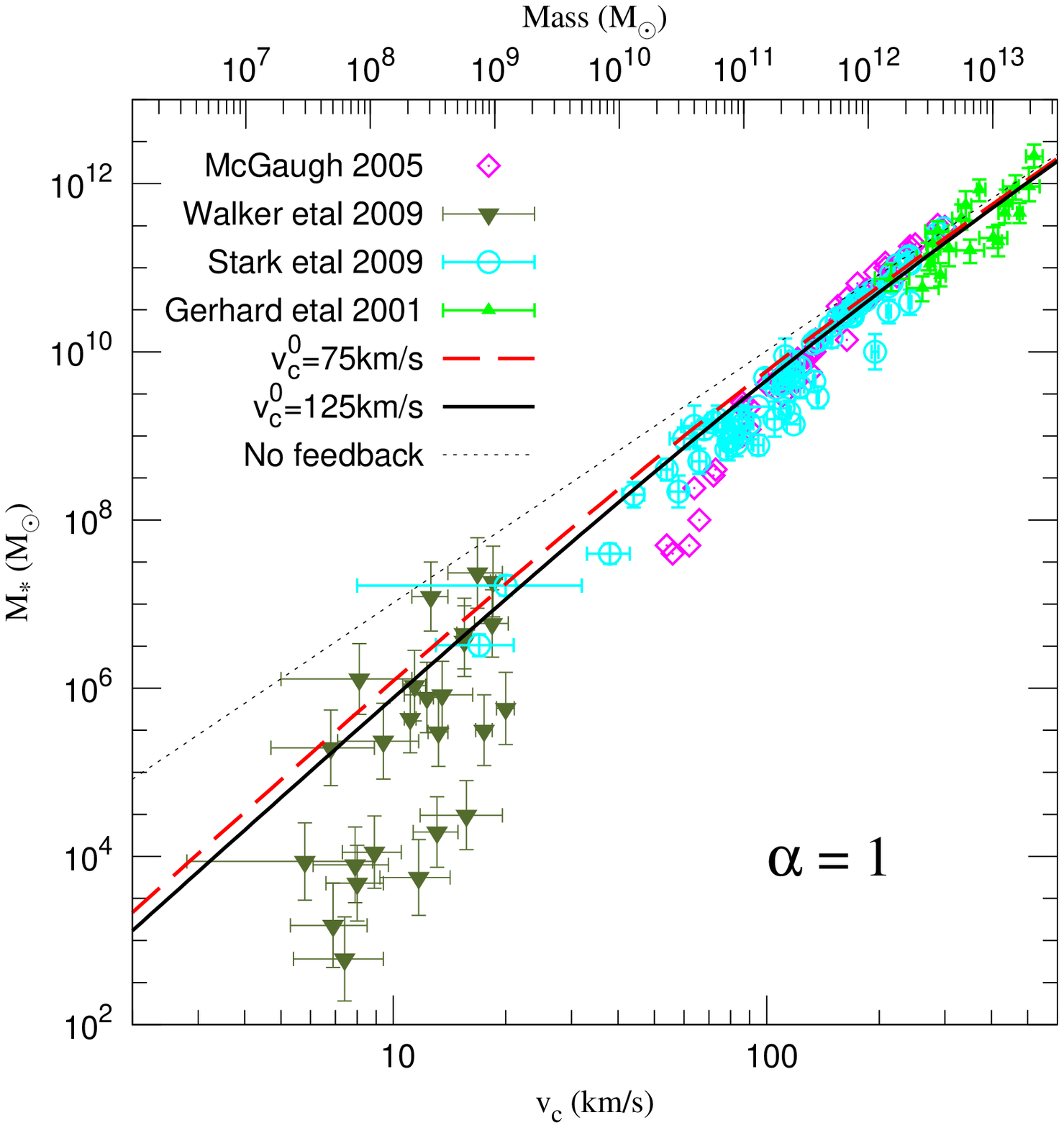,width=7.5cm,angle=-90.}}
\caption[]{Stellar mass as a function of circular velocity and mass of the galaxy.
The observed data points are taken from
McGaugh (2005) (magenta diamonds), Stark et al. (2009)(cyan circles),
Walker et al. (2009) (dark green reverse triangles) and Gerhard et al. (2001)
(green triangles).
The solid and dashed lines show our model prediction for
$v_c^0=125~$km/s and 75~km/s respectively with $f_*=0.4$. The left panel
is for $\alpha=2$ and right panel for $\alpha=1$. The dotted line in both
panels represents the no feedback scenario.
}
\label{fig_stellar_mass}
\end{figure*}

In Fig.~\ref{fig_stellar_mass} we show our model predictions of stellar masses
as a function of both circular velocity and mass of a galaxy along with the
observational data. The observed data points are taken from McGaugh (2005),
Stark et al. (2009), Walker et al. (2009) and  Gerhard et al. (2001). 
%
Note that in absence of any feedback, one expects $M_*=f_* (\Omega_b/\Omega_m)M$
which would be a straight line as shown by the dotted lines in
Fig.~\ref{fig_stellar_mass}. From observational data itself it is obvious
that small mass galaxies have very small amount of stars. At $M=10^8~M_\odot$
the suppression is almost two orders of magnitude compared to the universal
dark matter to baryon ratio whereas at $M=10^{12}~M_\odot$ the suppression is
almost zero. Thus star formation models that do not take account of feedback
fail to reproduce the observation. One must consider the negative feedback
of SNe on subsequent star formation in a consistent way like in our model. 
We show our model predictions for $\alpha = 2$ (left panel) i.e. energy
driven SNe feedback and $\alpha=1$ i.e. momentum driven SNe feedback.
The solid and dashed lines in both the panels are for $v_c^0=125$~km/s
and 75~km/s respectively. We choose such normalisation as they provide
a reasonable good fit to the data and also in Samui et al. (2010)
such a normalisation naturally arises for Cosmic ray driven outflows
(See Table~1 of Samui et al., 2010). We have assumed $f_*=0.4$ for all the
models. Note that with this choice of $f_*$, less than 20\% of baryon mass
is converted to star for $M \lesssim 10^{11}~M_\odot$.

It is clear from the figure that our model predictions match quite well with
the observed data. For high mass galaxies with $v_c\gtrsim 30$~km/s the errors
in the measurements as well as the scatter in the data are small and we obtain
a good fit to the observed data for both $\alpha=2$ and 1. The spread in the
data is within the statistical uncertainty of the model parameters. It is
interesting to note that the observed data sample consists of different types
of galaxies. While the Stark et al. (2009) galaxy samples are gas dominated
spiral galaxies, the McGaugh et al. data consists of both gas dominated as well
as gas poor star mass dominated galaxies. On the other hand, the Gerhard et al.
(2001) samples are compendium of the early type galaxies. Hence our model
reproduces the observed stellar mass fraction for different galaxy population
with masses $M\gtrsim 10^{10}~M_\odot$. For small mass dwarf galaxies with
$v_c\lesssim 30$~km/s the observational points are quite scatter although our
model predictions agree reasonably well especially for $\alpha=2$. For
$\alpha=1$ it seems that additional feedbacks other than the SNe feedback
are needed to suppress star formation even more in those galaxies in order
o explain majority of the data points. However in any model the spread in
the observational data is much more than the statistical uncertainty of the
model parameters. Interestingly these dwarf galaxies are mostly satellite
galaxies and they are likely to be affected by tidal striping and radiative
feedback from reionization that we discuss later while considering the
luminosity function of LBGs. Hence, it is not surprising
that our simple model of star formation including only the SNe feedback fails
to explain the spread in the observed data for dwarf galaxies.
Also note that for higher values of $\eta_w$ one may expect
the radiative cooling of the outflowing material to be important and
the hot thermal gas that drives the outflows cools efficiently making
the outflow a momentum driven one (see Samui et al. 2009 for details example
of such outflows). However, our models show that $\alpha=2$ is more
favourable even for low mass dwarf galaxies which points to the
fact that the outflows from dwarf galaxies are most likely not momentum
driven outflows. Therefore, one must consider some alternative
driving force for the outflows in these galaxies. In Samui et. al (2010)
we showed that cosmic rays can provide such alternative with correct scaling
of $\alpha=2$.

%
\begin{figure}
\centerline{\epsfig{figure=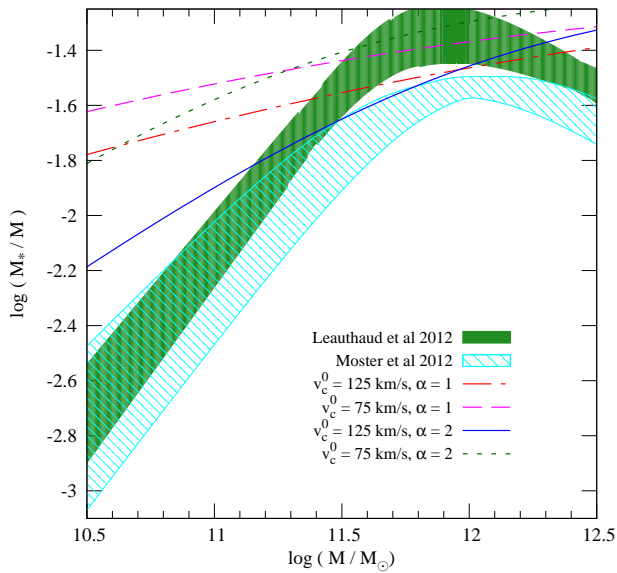,width=7.5cm,angle=-90.}}
\caption[]{The stellar to halo mass ratio. The green shaded and cyan hatched areas
are observationally derived SHM relation taken from Leauthaud et al. (2012) and Moster
et al. (2010) respectively. Different lines are predictions from our various models.
The blue solid and green dotted lines are for $\alpha = 2$ with
$v_c^0 = 125$ and 75~km/s respectively. The dotted dashed red
line and dashed magenta line are for $\alpha = 1$ with $v_c^0 = 125$
and 75~km/s respectively.}
\label{fig_SHM}
\end{figure}

In order to investigate this more we also consider the ratio of stellar mass
to halo mass (SHM) as a function of halo mass. In Fig.~\ref{fig_SHM} we show
our model predictions along with observationally derived SHM ratio by
Leauthaud et al. (2012) (green shaded region) and Moster et al. (2010)
(cyan hatched area) with measurement uncertainty. Leauthaud et al. used deep
COSMOS data along with halo occupation distribution model to constrain SHM
ratio. Moster et al. derived the SHM ratio from N-body simulation with recent
SDSS clustering data and galaxy galaxy lensing data. Even though there are
slight mismatch in their derived SHM ratio, our model predictions match well
considering the uncertainty in measurement. Here also we see that momentum
driven cases ($\alpha=1$) provide a poorer fit to the derived SHM ratio
compared to $\alpha=2$ models in low mass region. We wish to point out that the
decrease in SHM ratio for $M\gtrsim 10^{12}~M_\odot$ is due to AGN feedback
that we have not considered here. Also the observed galaxies have a limiting
stellar mass of $log(M_*/M_\odot)=8.7$ corresponds to $log(M/M_\odot) \sim 11$
that has been used to derived the SHM. Hence one should not compare our model
predictions with this particular observation beyond this mass limit.

\section{Mass-Metallicity relation}
\label{sec_mass_metal}
In this section we focus on the gas phase metallicity of the high redshift
galaxies in order to understand the observed mass-metallicity relation.
The metallicity of a galaxy in our model can be calculated as follows.
We assume that the amount of metals ejected by the supernova per unit of star
formation is $p$ and these metals are mixed with the ISM instantaneously.
Suppose at a given time the metallicity of the ISM is $Z$. If at that instance
$\delta M_*$ amount of stars are formed then total gas mass reduced from the
ISM is $\delta M_g = (1+\eta_w) \delta M_*$. The second term arises due to
the loss of gas through outflows. Note that we neglect the recycled
gas that is returned by the supernova into the ISM as it is very small.
For example, in a Salpeter IMF from $0.1-100~M_\odot$ the return
fraction is less than 15\%. Given $\delta M_*$ of star formation,
the amount of metal lost from the ISM is $Z(1+\eta_w)\delta M_*$.
However, supernova will produce $p \delta M_*$ metals. Hence total change of
metal mass ($M_h$) in the ISM is $$\delta M_h = [p-Z(1+\eta_w)]\delta M_*.$$
Recall that $Z=M_h/M_g$. Differentiating this and using above relations we obtain
\begin{eqnarray}
\delta Z & = & \frac{\delta M_h}{M_g} - \frac{M_h}{M_g^2} \delta M_g \nonumber \\
 &=& \left[\frac{p}{M_b-(1+\eta_w)M_*}\right] \delta M_* 
\end{eqnarray}
Integrating this with boundary condition that the initial metallicity
of the collapsed gas is $Z_0$, we obtain
\begin{equation}
Z=Z_0-\left(\frac{p}{1+\eta_w}\right)\ln\left[1-(1+\eta_w)\frac{M_*}{M_b}\right]
\label{eqn_mass_metal}
\end{equation}

\begin{figure}
\centerline{\epsfig{figure=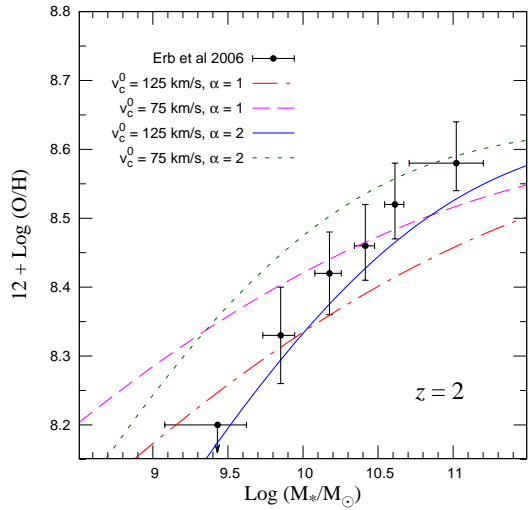,width=7.0cm,angle=-90.}}
\caption[]{The gas phase metallicity as a function of stellar mass. The blue
solid and green dotted lines are for $\alpha=2$ with $v_c^0 = 125$ and 75~km/s
respectively. The dotted dashed red line and dashed magenta line are for 
$\alpha=1$ with $v_c^0 = 125$ and 75~km/s respectively. The Observed data points
are taken from Erb et al., (2006).
}
\label{fig_mass_metal}
\end{figure}

In Fig.~\ref{fig_mass_metal} we show our model prediction for stellar mass
metallicity relation. Note that the stellar mass is obtained from
Eq.~\ref{eqn_m_total} and metallicity is obtained from
Eq.~\ref{eqn_mass_metal}. The observed data points are taken from Erb et al.
(2006) for the sample of galaxies at $z\sim 2$. We take $p=(0.6/50)~M_\odot$
which means one supernova will form per $50~M_\odot$ of star formation and it
will produce $0.6~M_\odot$ of Oxygen (Starburst99: Leitherer et al., 1999).
Further, we assume $Z_0=0$. It is clear from the figure that our models
reproduce the observed correlation very well. Especially, for the case of
$\alpha=2$ models, the two normalisation circular velocities namely $v_c^0=75$
and 125~km/s nicely bracket the observed correlation function. Predicted mass
metallicity correlation by $\alpha=1$ models is flatter than the observed
values. Therefore, this correlation also favours the $\alpha=2$ models compared
to $\alpha=1$ models which produce poorer fit to the observation. And we
conclude that the wind feedback is the main driver to determine the amount of
metals in galaxies.
%

\begin{figure*}
\centerline{\epsfig{figure=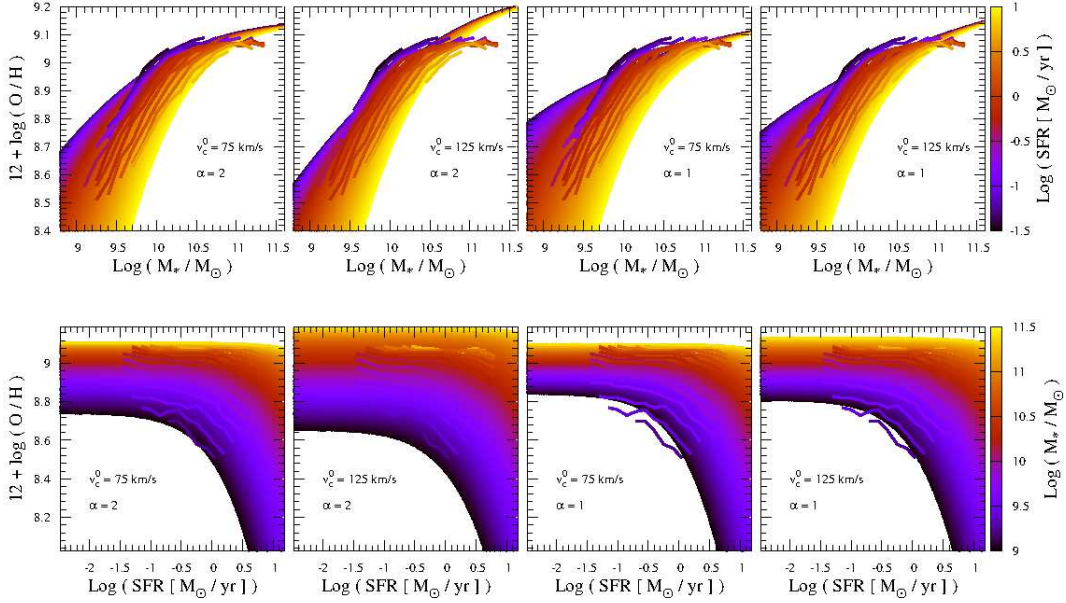,width=8cm,angle=-90}}
\caption[]{Projection of fundamental mass-metallicity-SFR relation. We show
our model predictions for the mass metallicity SFR relation by the colour
shaded area. The top panels show the metallicity as a function of stellar
mass. The color indicates the star formation rate. In bottom panels we show
the metallicity as a function of star formation rate. In these panels the
color indicates the stellar mass. In all panels the observed fundamental mass
metallicity SFR relation is shown by the same color coded solid curves taken
from Mannucci et al. (2010). 
}
\label{fig_FMR}
\end{figure*}

As we have discussed early, recent observations have pointed out that
the correlation of stellar mass and gas phase metallicity actually comes
from a projection of more fundamental 3-D relationship between stellar mass,
metallicity and star formation rate of galaxies. The offset seen in the
mass metallicity relation between local universe and high redshift universe
is due to selection bias of observing only higher star forming galaxies
at high redshift (Mannucci et al. 2010). Hence any model of galaxy formation
should be able to reproduce this fundamental 3-D relation. Indeed our models
do predict such a correlation between stellar mass, metallicity and SFR.
In Fig.~\ref{fig_FMR} we show both projections of this relationship
as predicted by our models along with the observational data taken from
Mannucci et al. (2010). The top panels show metallicity as a function of
stellar mass (the color shaded area) as obtained from all the four models
discussed above (the parameters of each model is indicated in each panel).
The color bar represent the star formation rate of the galaxy. In each panel
the color codded curves are the observational data with same color indicator
for the star formation rate. In the bottom panels we show the metallicity as
a function of star formation rate for a range of stellar masses. It is obvious
from the figure that our models produce the 3-D correlation observed in
galaxies. All our models produces the trend, i.e. for a given stellar mass,
higher star forming galaxies have lower metallicity compared to low star
forming counterparts or for a given star formation higher metallicity
is obtained for higher stellar mass. However, it is interesting to point out
that the model with $\alpha=2$ and $v_c^0=75$~km/s provides the best
fit with the observation. Note that in this case the model has to reproduce
both the projections simultaneously. Like, the model with $\alpha=2$
and $v_c^0=125$~km/s provide good fit with mass-metallicity relation
but does not reproduce spread in SFR-metallicity relation.
Both the models with $\alpha=1$ unable to explain the correctly
the observational data. We therefore conclude like in previous section
that the observations are well reproduced by $\alpha=2$ models
again indicating the importance of cosmic ray driven winds.
%

\section{Luminosity functions at high redshifts}
\label{sec_lf}
After successfully explaining the stellar and metal mass detected in galaxies,
we turn to the high redshift UV luminosity functions of LBGs. We broadly
follow Samui et al. (2007 \& 2009) to calculate high redshift luminosity
functions. Below we briefly describe our model.

We obtain luminosity evolution of a galaxy undergoing a burst of one unit of
star formation at a specific wavelength (say $\lambda = 1500$~\AA) from
population synthesis code ``Starburst99" (Leitherer et al., 1999) assuming some
initial mass function\footnote{We take a Salpeter initial mass function
in the mass range $1-100~M_\odot$ throughout this paper.} for the stars formed.
We convolve that with star formation rate of individual galaxy in our model
(i.e. Eq.~\ref{eqn_dM_dt_2}) to obtain the luminosity as a function of galaxy
age (See Eq.~6 of Samui et al., 2007 and also Fig.~1 there). Note that the
difference with Samui et al. (2007) model is that here we explicitly use the
SNe feedback to the star formation. Further, all the light produced inside
a galaxy by the stars do not reach to us due to presence of dust. A fraction,
$1/\eta$, of total luminosity of the galaxy can be observed from earth.

We assume each luminous galaxy is formed inside a virialised dark matter halo
provided the gas can cool and host star formation. The formation rate of dark
matter halos per unit volume, $N(M,z)$, at a given redshift is calculated by
taking the redshift derivative of Sheth-Tormen mass function (Sheth \& Tormen,
1999).
The Sheth-Tormen (ST) mass function provides a good fit to the numerical
simulation data (Brandbyge et al., 2010) and hence we use it to calculate
the formation rate of dark matter halos. Note that the time derivative of
mass function provides total change in the number of halos; not just formation
rate. One can extend Sasaki formalism (Sasaki, 1994) for Press-Schechter Mass
function in order to get formation rate of halos from ST mass function.
However, it is well know that naive extension of this formalism does not
work for other mass functions (Samui et al., 2009; Mitra et al, 2011).
Hence we use the simple time derivative to get the formation rate of halos
with the assumption that it closely follows the formation rate.
However, to show the effect of mass function on the predicted luminosity
functions, we also consider Press-Schechter (PS) mass function with Sasaki
formalism to calculate the formation rate of dark matter halos. In particular,
we want to see if the conclusion of Samui et al. (2009) is still valid
with the new feedback induced star formation model. The cumulative luminosity
function, $\Phi(>L, z )$ for luminosity $L$ at a given redshift $z$ can be
obtained from
\begin{equation}		
\Phi(>L, z ) = \int\limits_z^\infty \de z_c~ N(M,z_c) \Theta[L - L(M,z,z_c)].
\label{eqn_lf_cum}
\end{equation}
Here, $N(M,z_c)$ is the number of dark matter halos per unit volume between
mass $M$ and $M+dM$ and  collapsed between redshift $z_c$ and $z_c + dz_c$.
The Heaviside theta function $\Theta[L - L(M,z,z_c)]$ ensures that the
integral is contributed only by galaxies that are formed at redshift $z_c$
and having luminosity, $ L(M,z,z_c)$, greater than $L$ (after correcting for
dust reddening) at observe redshift $z$, with $z<z_c$. Taking derivative of
Eq.~\ref{eqn_lf_cum} with respect to $L$ and multiplying by the jacobi
$dM_{AB}/dL$ ($M_{AB}$ is the magnitude in AB system (Oke \& Gunn, 1983))
we get the luminosity functions, $\Phi(M_{AB},z)$, at a given redshift.

In our model we calculate the reionization history and radiative feedback to
the star formation in a self consistent way (for detail see Samui et al., 2007).
A galaxy forming in the neutral region can cool with the help of atomic cooling
if its virial temperature of the dark matter halo is greater than $10^4$~K.
Below this temperature a galaxy can cool and host star formation only in
presence of molecular hydrogen. In this paper we consider only the atomic
cooled halos as they are the main contributor to the luminosity functions
in the observable range. In the ionised region of universe, due to the increase
of Jeans mass, a galaxy can host star formation only if its virial velocity is
greater than 35~km/s. We assume a complete suppression of star formation for
$v_c\le35$~km/s and no suppression above $v_c=95$~km/s. For intermediate mass
range we adopt a linear fit from 1 to 0 (Broom \& Loeb, 2002; Benson et al.,
2002, Dijikstra et al., 2004). We also assume a suppression of star formation
in high mass halos by a factor of $[1+(M/10^{12}M_\odot)^3]^{-1}$ due to
possible feedback from AGN (Bower et al., 2006; Best et al., 2006).

\section{Constraining model parameters from observations}
\label{sec_fit}
In this section we show our model predictions for the UV luminosity functions
of LBGs and compare them with observations in the redshift range $1.5 \le z
\le 8$. Note that two free parameters of our model, the star formation
efficiency, $f_*$ and dust reddening correction factor, $\eta$ come as a
product like $f_*/\eta$ while calculating the luminosity of a galaxy and
hence the luminosity function at a given redshift (we have assumed $f_t=1$).
We use $\chi$-square minimization technique to fit observed luminosity
functions at different redshifts taking $f_*/\eta$ as a free parameter.
Taking this product ($f_*/\eta$) as independent of mass of the galaxy may
seem simplistic. However, note that both $f_*$ and $\eta$ can depend on the
mass of the galaxy (Bouwens et al., 2012). Since we can not disentangle one
from other while calculating the luminosity we do not consider mass dependent
$f_*$ and/or $\eta$ here. Further, $\eta$ measures the amount of reddening due
to dust. Observationally it has been found that the high redshift galaxies
exhibit smaller amount of dust compared to their low redshift counter part
by measuring the UV continuum slope (Hopkins \& Beacom, 2006; Bouwens et al.,
2012). From physical point of view, high redshift galaxies should have less
amount of metals and hence little dust. Thus, we expect that
$f_*/\eta$ should have different values at different redshift even if
the efficiency remains same and we try to constrain this by fitting
the observed UV luminosity function.

%
\begin{figure*}
\centerline{\epsfig{figure=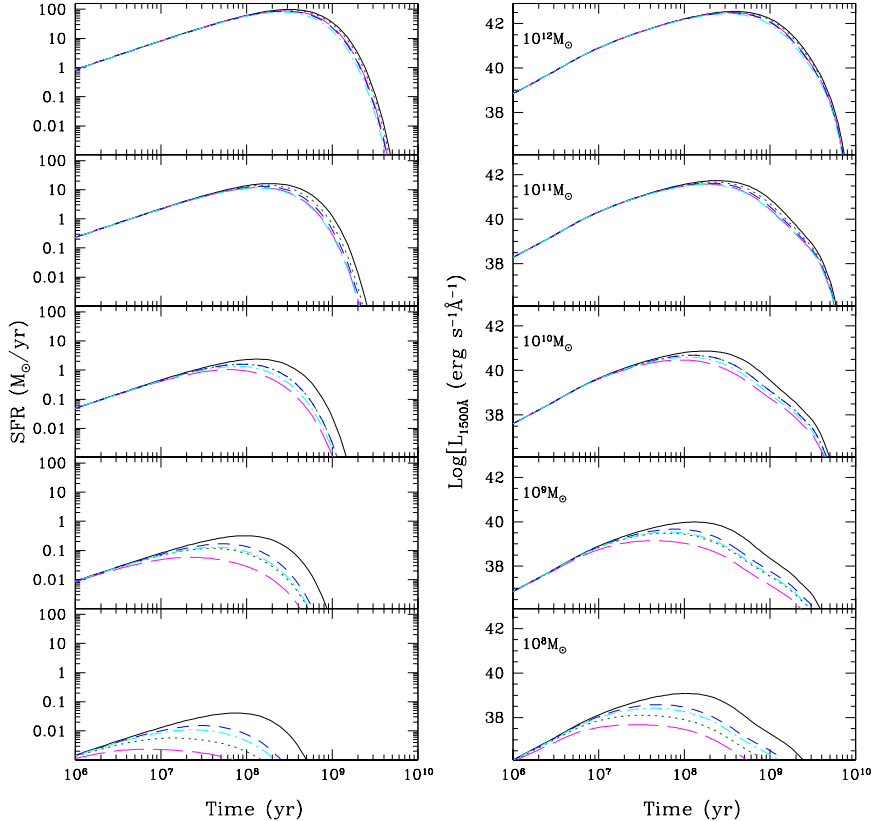,width=12.0cm,angle=0.}}
\caption[]{The evolution of star formation rate (left panels) and luminosity
at 1500~\AA~
(right panels) of individual galaxies of masses $10^{12}$, $10^{11}$, $10^{10}$,
 $10^{9}$ and $10^{8}~M_\odot$ from top to bottom. The dark green dotted lines
and magenta long dashed lines are for $\alpha=2$ with $v_c^0=75$ and
125~km/s respectively. The blue short dashed curves and cyan dotted dashed
curves are for $\alpha=1$ with $v_c^0=75$ and 125~km/s respectively.
In each panel we show, for comparison, the without feedback model (i.e.
$\eta_w=0$) with solid black lines.  
}
\label{fig_sfr_lf}
\end{figure*}

Before showing the luminosity functions at a given redshift, we show the star
formation rate as obtained from Eq.~\ref{eqn_dM_dt_2} and resulting luminosity
evolution of a single galaxy as a function of galaxy age. In
Fig.~\ref{fig_sfr_lf}, we show star formation rate (SFR) (left panels) and UV
luminosity at 1500~\AA~  (right panels) for galaxies with masses $10^{12}$,
$10^{11}$, $10^{10}$,  $10^{9}$ and $10^{8}~M_\odot$ from top to bottom
assuming collapsed redshifts of $z_c=4$, 6, 8, 10 and 12 respectively. Note
that $z_c$ fixes the dynamical time scale $\tau$ (Barkana \& Loeb, 2001)
and we choose $z_c$ such that the halo can collapse from a $3\sigma$ density
fluctuation at $z_c$. We also take $f_t=1$ for the rest of the paper. Note that
changing $f_t$ would change the duration of star formation. Larger $f_t$
means a shorter duration of star formation representing a burst mode of
star formation and smaller $f_t$ corresponds to a prolong star formation
scenario in the galaxy.
In Fig.~\ref{fig_sfr_lf} we show models with $\alpha=2$ \& $v_c^0=125$~km/s
(dark green dotted curves), $\alpha=2$ \& $v_c^0=75$~km/s (magenta long dashed
curves), $\alpha=1$ \& $v_c^0=125$~km/s (blue short dashed curves), and
$\alpha=1$ \& $v_c^0=75$~km/s (cyan dotted dashed curves). For comparison
we also show the without feedback model ($\eta_w=0$) with black solid line.

Fig.~\ref{fig_sfr_lf} nicely demonstrates that our SNe feedback models affect
the star formation most in the low mass galaxies as $\eta_w$ is higher in those
galaxies. The maximum suppression of star formation  happens for models with 
$\alpha=2$ and $v_c^0=125$~km/s and the effect is minimum for models with
$\alpha=1$ and $v_c^0=75$~km/s. For $10^8~M_\odot$ galaxy the suppression is
almost two orders of magnitude in star formation rate that results two
orders of magnitude decrease in the maximum luminosity compare to no
feedback model (bottom panels of Fig.~\ref{fig_sfr_lf}). Note that peak of the
star formation hence the luminosity also happens early in time with
$t_{\rm peak}=(\tau/\eta_w)\ln(1+\eta_w)$. The suppression in star formation
decreases with increasing galaxy mass; we see one order of magnitude
suppression in peak star formation rate and also in maximum luminosity for
galaxy of $M=10^{10}~M_\odot$ (middle panels of Fig.~\ref{fig_sfr_lf}).
For even higher mass galaxies SNe feedback has very little effect on the star
formation and hence on the luminosity of the galaxy. One hardly notices the
difference (less than factor 2) in star formation rate/luminosity
for various feedback models with no feedback model for $10^{12}~M_\odot$ galaxy
(top panels of Fig.~\ref{fig_sfr_lf}). Interestingly such kind of feedback effect
due to the SNe on the star formation history of individual galaxies are also
seen in full hydrodynamic simulations (Scannapieco et al., 2006)

\begin{figure*}
\centerline{\epsfig{figure=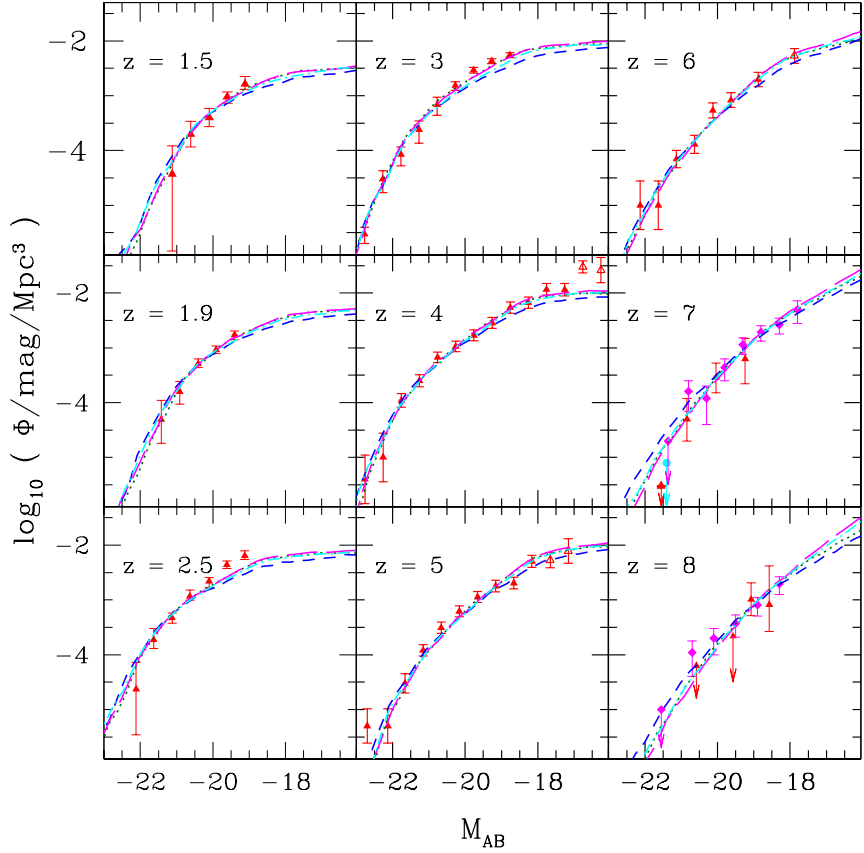,width=12.0cm,angle=0.}}
\caption[]{The observed luminosity functions for $1.5\le z \le 8$ along
with our model predictions for Sheth-Tormen mass function.
The best fit values of $f_*/\eta$ are tabulated
in Table~\ref{tab_fit}. We show our model predictions for $\alpha=2$ with
$v_c^0=75$~km/s (dark green dotted curves) and $v_c^0=125$~km/s (blue short
dashed curves) and $\alpha=1$ with $v_c^0=75$~km/s (magenta long
dashed curves) and $v_c^0=125$~km/s (cyan dotted dashed curves).
The observed data points are taken from Oesch et al. (2010)
(red triangles at $z=1.5$, 1.9 and 2.5),
Reddy \& Steidel (2009) (red triangles at $z=3$),
Bouwens et al. (2007) (red triangles at $z=4$, 5 and 6), Bouwens et al. (2008)
(red triangles at $z=7$), Mannucci et al. (2007) (cyan bullet at $z=7$)
and Bouwens et al. (2011) (magenta diamonds at $z=7$ and 8).
}
\label{fig_lf_fit}
\end{figure*}

\begin{figure*}
\centerline{\epsfig{figure=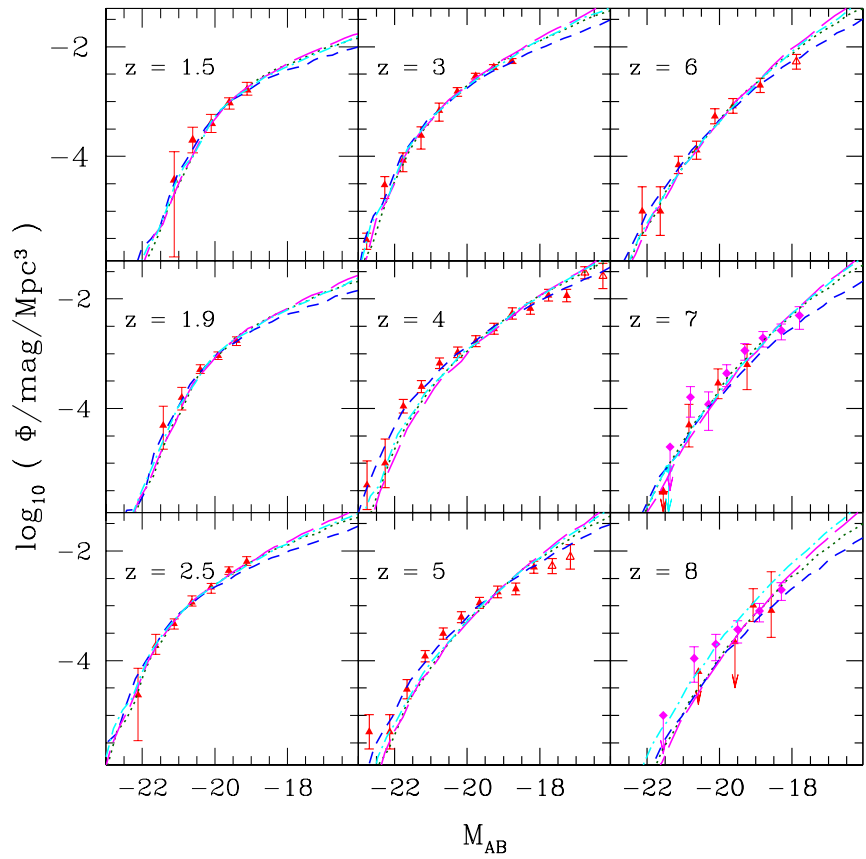,width=12.0cm,angle=0.}}
\caption[]{Same as Fig.~\ref{fig_lf_fit} but for Press-Schechter mass function
with Sasaki formalism.
}
\label{fig_lf_fit_PS}
\end{figure*}

Hence our SNe feedback model affects and regulates the star
formation/luminosity of galaxies differentially, low mass galaxies get affected
most and high mass galaxies least. Such a feedback would change the slope of
the luminosity functions at different redshifts making them flatter. Indeed we
see such flattening in the observational data of UV luminosity functions of
LBGs. In Figs.~\ref{fig_lf_fit} and \ref{fig_lf_fit_PS}, we show our model
predictions of luminosity functions in the redshift range $1.5 \le z \le 8$
along with the observational data for ST and PS mass function respectively.
The observed data points are taken from Oesch et al. (2010) ($z=1.5$, 1.9 and
2.5),  Reddy \& Steidel (2009) ($z=3$), Bouwens et al. (2007) ($z=4$, 5 and 6)
and Mannucci et al. (2007), Bouwens et al. (2008) and Bouwens et al. (2011)
for $z=7$ and 8. Note that the open triangles at $z=4$, 5 and 6 suffer from
incompleteness problem (see Bouwens et al., 2007) and hence we do not consider
them while fitting. The fitted values of $f_*/\eta$ along with the best fit
$\chi$-square per degree of freedom (dof) are tabulated in Table~\ref{tab_fit}.
Note that all our models predict the reionization histories that are well
within the available observational constraints. In particular, for all our
models, we obtain redshift of reionization greater than 6 as inferred from
the observations of lyman-alpha forest in distant quasars (Fan et al., 2006)
and electron scattering optical depth within one sigma of WMAP 7yrs
results ($\tau_e=0.087 \pm 0.014$) (Komatsu et al., 2011).
It is important to note that the feedback induced star formation model
reduces the star formation severely in low mass galaxies. These galaxies
are the main sources of the UV photon that causes the reionization
of the IGM and lower the star formation lesser the UV photon production.
However, the fraction of UV photon that can escape from a galaxy and cause
the ionisation of the IGM is poorly constraint especially for low mass
galaxies. We observe that using a escape fraction of 0.2 leads to
a similar reionization history (and electron scattering optical depth
to the reionization) for the feedback induced star formation models
compared to the no feedback model with escape fraction of 0.1. This
value of escape fraction of UV photon is in good agreement with
recent numerical results (Wise \& Cen 2009; Pawlik et al. 2009;
Yajima et al. 2009) as well as observations (Bouwens et al. 2010).
Thus, our feedback models also produce the reionization history
with reasonable physical parameters.

%
%

%
%
%
%

\begin{sidewaystable}
\begin{center}
\hskip -1cm%
\begin{tabular} {c|c|c|c|c|c|c|c|c} \hline \hline
$z$	&\multicolumn{2}{|c|}{$\alpha=2$, $v_c=125$~km/s}	&\multicolumn{2}{|c|}{$\alpha=2$, $v_c=75$~km/s} &\multicolumn{2}{|c|}{$\alpha=1$, $v_c=125$~km/s} &\multicolumn{2}{|c|}{$\alpha=1$, $v_c=75$~km/s} \\ \cline{2-9}
&	$f_*/\eta$	&	$\chi^2$/dof &	$f_*/\eta$	&	$\chi^2$/dof & 	$f_*/\eta$	&	$\chi^2$/dof & 	$f_*/\eta$	&	$\chi^2$/dof \\ \hline 
 \multicolumn{9}{c}{ST derivative}
\\ \hline

1.5	&	0.048 $ \pm $ 0.003  &	0.81 &	0.037 $ \pm $ 0.002  &	0.28 &	0.053 $ \pm $ 0.003  &	0.56 &	0.044 $ \pm $ 0.003  &	0.40 \\ \hline
1.9	&	0.049 $ \pm $ 0.002  &	0.89 &	0.038 $ \pm $ 0.001  &	0.24 &	0.054 $ \pm $ 0.002  &	0.49 &	0.045 $ \pm $ 0.001  &	0.22 \\ \hline
2.5	&	0.079 $ \pm $ 0.003  &	4.12 &	0.064 $ \pm $ 0.002  &	2.44 &	0.087 $ \pm $ 0.003  &	3.05 &	0.074 $ \pm $ 0.002  &	2.46 \\ \hline
3	&	0.059 $ \pm $ 0.001  &	6.40 &	0.056 $ \pm $ 0.001  &	1.51 &	0.067 $ \pm $ 0.001  &	4.13 &	0.062 $ \pm $ 0.001  &	1.80 \\ \hline
4	&	0.062 $ \pm $ 0.002  &	1.11 &	0.052 $ \pm $ 0.001  &	0.40 &	0.068 $ \pm $ 0.002  &	0.46 &	0.060 $ \pm $ 0.001  &	0.38 \\ \hline
5	&	0.050 $ \pm $ 0.001  &	0.98 &	0.041 $ \pm $ 0.001  &	1.36 &	0.052 $ \pm $ 0.001  &	1.14 &	0.045 $ \pm $ 0.001  &	1.51 \\ \hline
6	&	0.062 $ \pm $ 0.002  &	0.68 &	0.052 $ \pm $ 0.002  &	0.55 &	0.066 $ \pm $ 0.002  &	0.56 &	0.057 $ \pm $ 0.002  &	0.51 \\ \hline
7	&	0.120 $ \pm $ 0.006  &	0.38 &	0.083 $ \pm $ 0.004  &	0.38 &	0.086 $ \pm $ 0.004  &	0.40 &	0.072 $ \pm $ 0.003  &	0.50 \\ \hline
8	&	0.114 $ \pm $ 0.007  &	0.26 &	0.080 $ \pm $ 0.004  &	0.45 &	0.099 $ \pm $ 0.005  &	0.55 &	0.083 $ \pm $ 0.004  &	0.65 \\ \hline
\multicolumn{9}{c}{PS Sasaki} \\
\hline

1.5	&	0.029 $ \pm $ 0.001  &	0.12 &	0.023 $ \pm $ 0.001  &	0.25 &	0.031 $ \pm $ 0.001  &	0.21 &	0.026 $ \pm $ 0.001  &	0.24 \\ \hline
1.9	&	0.034 $ \pm $ 0.001  &	0.26 &	0.027 $ \pm $ 0.001  &	0.64 &	0.037 $ \pm $ 0.001  &	0.68 &	0.031 $ \pm $ 0.001  &	0.62 \\ \hline
2.5	&	0.067 $ \pm $ 0.002  &	1.45 &	0.054 $ \pm $ 0.001  &	0.37 &	0.074 $ \pm $ 0.002  &	0.50 &	0.063 $ \pm $ 0.002  &	0.28 \\ \hline
3	&	0.061 $ \pm $ 0.001  &	1.24 &	0.050 $ \pm $ 0.001  &	1.55 &	0.065 $ \pm $ 0.001  &	0.78 &	0.057 $ \pm $ 0.001  &	1.57 \\ \hline
4	&	0.074 $ \pm $ 0.001  &	1.02 &	0.051 $ \pm $ 0.001  &	4.55 &	0.066 $ \pm $ 0.001  &	4.58 &	0.052 $ \pm $ 0.001  &	6.76 \\ \hline
5	&	0.076 $ \pm $ 0.001  &	2.20 &	0.052 $ \pm $ 0.001  &	5.95 &	0.069 $ \pm $ 0.001  &	5.18 &	0.058 $ \pm $ 0.001  &	6.44 \\ \hline
6	&	0.115 $ \pm $ 0.004  &	0.63 &	0.090 $ \pm $ 0.003  &	0.71 &	0.110 $ \pm $ 0.003  &	0.77 &	0.096 $ \pm $ 0.002  &	0.80 \\ \hline
7	&	0.179 $ \pm $ 0.008  &	0.32 &	0.113 $ \pm $ 0.004  &	0.69 &	0.134 $ \pm $ 0.006  &	1.02 &	0.110 $ \pm $ 0.004  &	1.21 \\ \hline
8	&	0.248 $ \pm $ 0.014  &	0.32 &	0.153 $ \pm $ 0.007  &	0.80 &	0.185 $ \pm $ 0.009  &	0.86 &	0.150 $ \pm $ 0.007  &	1.06 \\ \hline
\hline

\end{tabular}
\caption{The best fit values of $f_*/\eta$ for all our models along with the
best fit $\chi$-square per degree of freedom for $1.5 \le z \le 8$. We quote
$1-\sigma$ error in the measurements of $f_*/\eta$.}
\label{tab_fit}
\end{center}
\end{sidewaystable}
%

We first consider the ST mass function. It is clear from the
Fig.~\ref{fig_lf_fit} and also from Table~\ref{tab_fit} that our models
reasonably explain the UV luminosity functions of LBGs in a vast redshift
range of $1.5\le z \le 8$. The flattening observed in the low end of
luminosity functions ($M_{AB}\gtrsim 18$) particularly in the redshift range of
$4\le z\le 6$ are well explained by the SNe feedback and the radiative feedback
that we assume. Except for $z=2.5$ and 3 the best fit $\chi$-square per
degree of freedom is very close to unity or less than unity. This demonstrates
how well our model predictions match with observations.
In general, for all our models, values of $f_*/\eta$ show an increasing
trend with redshift with some exceptions at $z=2.5$ and $z=5$
that we discuss later.
This is in accord with the concordance model of structure formation.
If one assumes a fixed $f_*=0.4$ which provides a reasonable fit to
the observational data for the stellar mass in galaxies (See
Fig.~\ref{fig_stellar_mass}) one sees that the value of $1/\eta$
increases with increasing redshift. This means in the past the
amount of dust present in a galaxy was less on average.
As the universe became older the amount of metals in galaxies
increased making them more dusty and hence more difficult to detect.
Also the mean values of $\eta$ obtained assuming $f_*=0.4$
is in good agreement with the observation of Reddy et al. (2012). They found
a mean value of $\eta=5.2$ at $z\sim 2$.

Further we notice that the models with $v_c^0=75$~km/s provide a better fit
to the observational data compare to the models with $v_c^0=125$~km/s for both
$\alpha=2$ and 1 in all redshifts except for $z=5$ and 8. In particular,
for $z\le 4$ the models with $v_c^0=75$~km/s fit the faint end slope
of the luminosity functions better than models with $v_c^0=125$~km/s.
For $z\ge 5$ both the models over fit the observational data as
best fit values of $\chi^2$ per degree of freedom are less than unity and
hence it can not identify the best fit model. At $z=2.5$ none
of our models provide a good fit to the observational data. The bright
end of the luminosity function is well explained by our models. However,
they fail to reproduce the last two data points at the low luminosity end.
Interestingly we found the faint end slope of the observed luminosity function
at $z=2.5$ is quite steep compared to $z=3$ luminosity function. It is very
unlikely to change the slope within such a short time. We expect future
improve observations would clarify this dispute.
This also leads to a higher value of $f_*/\eta$ at $z=2.5$
compared to values that are obtained for $z \ge 3$.

The observed luminosity function at $z=3$ clearly favours the models with
$v_c^0=75$~km/s for both $\alpha=2$ and 1. The models with $v_c^0=125$~km/s
fail to reproduce the faint end slope of the observed luminosity function.
Such a signature is also seen in case of $z=1.5$ and $z=1.9$. The feedback
in those models (i.e. models with $v_c^0=125$~km/s) are very strong to
suppress the star formation in low mass galaxies and hence fail to produce
enough number of low luminosity galaxy to explain the observed luminosity
functions at $z\le 3$.

At $z=4$ we see that our models provide a reasonable fit to the observed
data. The model with $\alpha=2$ and $v_c^0=125$~km/s does not fit the last four
observational points in the low luminosity end and hence has maximum value
for the best fit $\chi^2$. However, none of our models able to fit the last
two data points in the low luminosity end (the open triangles). As already
mentioned above these two data points suffer major uncertainty in measurements
and we do not consider them while fitting. If future observations confirm these
data points then we have to change our model in order to explain
the low luminosity end of the luminosity function at $z=4$.

All our models well explain the observed luminosity function at $z=5$.
However, we see from Table~\ref{tab_fit} that the best fit values of $\chi^2$
per dof at this redshift are quite higher than unity except for model with
$\alpha=2$ and $v_c^0=125$~km/s for which best fit $\chi^2$/dof~$=0.98$. We
observe that the data point at $M_{AB}=-18.66$ has unusually low value of
$\Phi(M_{AB})$ compare to its neighbouring data points and contribute
significantly to the best fit $\chi^2$ values. It also leads to a lower value
of $f_*/\eta$ compared to $z=4$.  Ignoring this data point all our models fit
the observed luminosity function with best fit $\chi^2$ per dof values close
to unity.

For redshift $z\ge 6$ our model predictions match very well with the observed
data. Note that the errors in the measurements are very high especially
at $z=7$ and 8. Thus all models provide good fit to the data with best fit
$\chi^2$ per degree of freedom much less than unity. We expect that future
observations would reduce the error bars and extend the measurements
to even lower luminosity allowing us to put further constraints on star
formation models. In passing we note that we can not distinguish two feedback
models namely energy driven/cosmic ray driven SNe winds (i.e. $\alpha=2$) and
momentum driven SNe winds ($\alpha=1$) with the present observational data of
UV luminosity functions of LBGs at various redshifts. The reason behind this
is that the two models differ maximum in low mass dwarf galaxies where the
$\eta_w$ can be different by more than factor 10. It was shown in Samui et al.
(2007) that the present observable range of luminosity functions are
contributed mostly by the galaxies with mass $M\gtrsim 10^{10}~M_\odot$.
At this mass the $\eta_w$ for models with energy driven SNe wind is larger
only by factor 3 (2) in case of $v_c^0=125$~km/s ($v_c^0=75$~km/s)
compared to models with momentum driven wind and hence can not be distinguished
by the present observational data due to the large errors in the measurements.

We now turn to the luminosity function as obtained using PS
halo mass function and compare with ST mass function.
Fig.~\ref{fig_lf_fit_PS} shows the predicted luminosity functions
and the fitted parameters are tabulated in Table~\ref{tab_fit}.
It is obvious from the figure that PS mass function also produces
the shape and evolution of luminosity function correctly.
However, Table.~\ref{tab_fit} shows that in most of the redshift
we considered (i.e. except $z=2.5$ \& 3),
PS mass function provide a poorer fit compared to ST mass function.
Thus the flattening of luminosity functions at low luminosity end
can only be explain due to SNe feedback and can not be
explain with different form of the halo mass function. Further,
our previous conclusions regarding the halo mass function remain
same i.e. ST mass function provides a better understanding of high redshift
luminosity functions compared to PS mass function.

\section{Discussions and Conclusions} 
\label{sec_cd}
We have built improved semi-analytical models of high redshift galaxy formation
where  star formation is regulated by the feedback due to SNe driven outflows.
We consider two models of feedback; one in which the outflows are driven
by the thermal energy of the SNe remnants and in second the hot gas loses its
thermal energy due to radiation and the momentum of the gas helps in driving
the outflow. The effect of SNe feedback to the star formation is calculated in
a self-regulated manner. We derive an analytical form for the star formation
rate in a galaxy that is regulated by the SNe feedback. Given the star
formation rate in individual galaxy we calculate high redshift galaxy
luminosity functions.

Our feedback induced star formation models are successful in explaining
the observed stellar mass in galaxies of different types
with mass range $10^{10}~M_\odot\lesssim M \lesssim 10^{13}~M_\odot$.
For low mass dwarf galaxies our simple model of star formation
including SNe feedback produces the trend in correlation
between the observed stellar mass and circular velocity but fails
to explain the spread in it. Other feedback mechanisms are expected
to operate on such small mass galaxies in order to understand the observations.
Our models also produces the observed ratio of stellar mass to
halo mass as a function of halo masses as obtained from recent
SDSS data. Further, our models emphasize the importance of alternative driver
of galactic outflows such as cosmic rays. 
The amount of metals detected in high redshift galaxies is well
estimated by our models which demonstrates that the metal mass
in a galaxy is vastly determined by the outflows.
The 3-D correlation between gas phase metallicity, stellar
mass and star formation rate is well explained by our
feedback dominated star formation model showing the importance
of SNe feedback in building this observed relationship in galaxies.

The observed UV luminosity functions of LBGs at $1.5 \le z \le 8$
are well explained with our feedback models. Especially the flattening
observed at low end of the luminosity functions arises naturally
with our feedback model. In absence of such feedback one would
produce more low mass galaxies and fail to reproduce the observed
data points. The models with $v_c^0=75$~km/s provide
a better fit to the observational data compared to $v_c^0=125$~km/s
especially at low redshift (the characteristic circular velocity,
$v_c^0$, fixes the normalisation of our feedback model).
This implies that the supernova feedback affects the star formation
in galaxies more compare to ionisation feedback for high mass
galaxies. For galaxies with $v_c\gtrsim 90$~km/s one expects
to have no effect of ionization feedback to the star formation
where as SNe feedback reduces the star formation by 40\% in the
same galaxies (taking $v_c^0=75$~km/s). The present observational data
of high redshift luminosity functions is not good enough to distinguish
a thermal energy driven outflow from a momentum driven outflow.
We need to reduce the error in the measurements and extend the
data to even lower luminosity end where these two models
behave differently. Moreover, the flattening of the luminosity functions
can not be explained as the difference between the form of halo
mass function that one uses to calculate the formation
rate of halos (like PS or ST mass function). It can
only be understood due to SNe feedback.

Hence, our feedback regulated star formation model can explain
a vast range of observations of high redshift universe.
Having such kind of simple analytical form for physically motivated
star formation of individual galaxy is very helpful for determining
the star formation rate of high redshift galaxies by fitting spectral
energy distribution (SED) with several broad band photometry.
In such case the star formation history is very important
to determine several physical quantities such as 
the star formation rate, stellar mass, age of the
galaxy etc. In general one considers a constant star formation
model or an exponentially decaying one in such fits.
However, Reddy et al. (2012) showed that high redshift
galaxies ($1.5\le z\le 2.6$) are more expected to have a rising
star formation history and very young age when they are detected.
Our successful star formation models actually predict such
kind of rising star formation history at the young age
and then gradually decreasing star formation history
after a fraction of dynamical time. We expect that using our model
in such SED fitting would lead to a better fit to the observational
data and also provide more clearer picture of high redshift galaxies.

Since SNe feedback is one of the main improvement of our model over
past, it is important to investigate its effect on the global
metal pollution. Especially by the fact that in low mass
galaxies that dominate in the metal production at high redshift
should have a cosmic ray pressure driven outflows, one should
see how far this model can explain the global metal budget.
We hope to address that in future.

\section*{acknowledgements}
The author thanks Kandaswamy Subramanian and Raghunathan Srianand
for very useful discussions. The author also thanks anonymous
referee for his/her useful comments that helped a lot to
improve the quality of the paper.

\end{document}